\documentclass[preprint,amsmath,amssymb,showkeys,nofootinbib]{revtex4}

\newcommand{\mbf}[1]{\mbox{\boldmath $#1$}}

\def\0{\mbox{\mbf 0}}

\def\h{\mbox{\mbf h}}

\def\x{\mbox{\mbf x}}

\def\p{\mbox{\mbf p}}
\def\dotp{\dot{\mbox{\mbf p}}}
\def\ddotp{\ddot{\mbox{\mbf p}}}
\def\q{\mbox{\mbf q}}
\def\n{\mbox{\mbf n}}

\def\RR{\prescript{(3)\!\!} {}R}

\def\BibTeX{{\rm B\kern-.05em{\sc i\kern-.025em b}\kern-.08em
    T\kern-.1667em\lower.7ex\hbox{E}\kern-.125emX}}

\usepackage{graphicx}

\usepackage{dcolumn}

\usepackage{bm}

\usepackage{mathtools}

\begin{document}


\title{Tunneling Through Bridges:  Bohmian Non-locality From Higher-Derivative Gravity}

\author{Gregory S. Duane}


\affiliation{
Geophysics Institute, University of Bergen, Postboks 7803, 5020 Bergen, Norway and \\ Dept. of Atmospheric and Oceanic Sciences, University of Colorado, Boulder, CO 80309, USA }
%
%
\begin{abstract} A classical origin for the Bohmian quantum potential, as that potential term arises in the quantum mechanical treatment of black holes and Einstein-Rosen (ER) bridges, can be based on 4th-order extensions of Einstein's equations.   The required 4th-order extension of general relativity is given by adding quadratic curvature terms with coefficients that maintain a fixed ratio, as their magnitudes approach zero, with classical general relativity as a singular limit.   If entangled particles are connected by a Planck-width ER bridge, as conjectured by Maldacena and Susskind, then a connection by a traversable Planck-scale wormhole, allowed in 4th-order gravity, describes such entanglement in the ontological interpretation.  It is hypothesized that higher-derivative gravity can account for the nonlocal part of the quantum potential generally.
\end{abstract}



\maketitle


%

%

\section{Motivation and Introduction} 

Non-locality is a prominent feature of quantum mechanics both in the standard, probabilistic Copenhagen interpretation and in
the deterministic re-interpretation by Bohm \cite{Bohm52}. In the former case, the spin of each member of an EPR pair,
along any measurement axis, is determined at the time of measurement of either member's spin along that axis, despite 
any space-like separation, however large, between the measurement apparatus and the remote spin.  Bell's Theorem \cite{Bell}
precludes any understanding of this result in terms of spin values that exist at the time of creation of the pair, independently of 
observation.  In Bohm's interpretation, the effect of measurement on the remote spin is taken to be a real physical effect, mediated 
by a quantum potential. 

With either interpretation, one is left to ask if any sort of generalization of the classical, and relativistic,  notion of locality can be articulated
that does not depend on the full apparatus of standard quantum theory.  If observations trigger the collapse of a globally defined wave packet,
as in the Copenhagen interpretation, can one be more specific about the aspect of the observational process that is responsible for the collapse?
If one adopts Bohm's quantum potential interpretation, can such a non-local potential be linked to interactions that are familiar in classical physics?
One must account for the impossibility of non-local transmission of information either through the alleged effect of observation or through the quantum potential.

A picture of quantum non-locality emerges from the Maldacena-Susskind ``ER=EPR" conjecture that was originally put forward
to resolve the black-hole information loss paradox \cite{EPEPR}.  We do not review the original motivation for ER=EPR, which involved entangled macroscopic
black holes, but simply note that what emerged was a suggestion that any pair of entangled particles is connected by an Einstein-Rosen (ER) bridge of Planck-scale width. Non-locality,
in the proposed view, is mediated by ER bridges, which intimately link space-like separated regions. Indeed, the members of an EPR pair have zero separation, 
as computed along a path through the bridge and thus are, in effect, the same particle. An Einstein-Rosen bridge is an unfolding of a black hole
in a multiply-connected space time topology, so information that enters the bridge is not transmitted for the same reason that it cannot escape a black hole.
That would explain why the non-local connections cannot be used for signaling.

It is not clear how Einstein-Rosen bridges, while offering a more specific and ontological picture, could lead to the kind of non-locality that would be induced 
by observations Copenhagen-style.  But it is possible to compare the physical effects of such bridges with those of the quantum potential in the Bohm
interpretation.  Effects like those of the quantum potential would require that the bridges be traversable, in order to resolve the paradox posed by the violation of the Bell inequality, which  is formulated in terms of measurement outcomes outside the bridge. Information must leave the bridge  for the outcome of measuring the spin of one member of an EPR pair to feel the orientation of the measurement apparatus for its partner, as required by Bell's Theorem.  Classical ER bridges are not traversable. However we note that
just as quantum black holes emit radiation, quantum ER bridges are crudely traversable. That is, particles can tunnel out of the black hole or through the bridge.
 There are severe restrictions on what can be re-emitted
in one case, or transmitted, in the other.  Thus the consequences of  tunneling through an Einstein-Rosen bridge are not to be feared and no new physics is required  that is beyond general relativity or quantum mechanics.

In the Bohm interpretation, as extended to field theory, tunneling through quantum ER bridges should be describable in terms of definite trajectories, giving a concrete, objective account of non-locality that is not possible in the standard quantum picture involving only probabilities and expectation values. New trajectories would arise because of
a quantum potential term that is added to the equations of classical general relativity.   In this interpretation,  tunneling mediated by a quantum potential cannot be used for superluminal signaling, because the statistics of an ensemble over all ``hidden variable" values are not affected, and such hidden variables cannot be controlled for an individual particle, though the physics is perfectly deterministic  (Bohm and Hiley  \cite{BohmBook} Sections 7.2, 12.6).    An extension of general relativity could thus safely account for quantum entanglement. Here, we assume that general relativity describes phenomena down to well below the Planck scale $L_P$.  In bootstrap fashion, the quantum behavior of  black holes and ER bridges, re-interpreted classically a la Bohm, would provide an explanation for 
quantum non-locality generally.  But what kind of geometrodynamics would give rise to the quantum potential required in the specific case of an ER bridge? The purpose of this Letter is to examine an extension of Einstein's equations that will account for the extra term and will thus allow us to
interpret the quantum ER transmission effect in terms of definite trajectories of particles and fields.

The proposed ontological account of quantum non-locality tends to resolve the contradiction between quantum mechanics and general relativity in a way that gives primacy to the latter.
In this regard, the proposal can be compared to Penrose's suggestion that the collapse of the wavefunction is a gravitational effect \cite{PenroseENM}.  After reviewing both the quantum potential interpretation of black hole radiation and classical higher derivative gravity in the next section, we compare the two in the following section, showing that 4th-order gravity is enough to simulate the quantum potential. The main result is that only infinitesimal 4th-order corrections are required, so that the standard second-order theory is a singular limit.  Implications for a possible re-interpretation of quantum theory  generally are discussed in the final section.

 \begin{figure}
a)\resizebox{.37\textwidth}{!}{\includegraphics{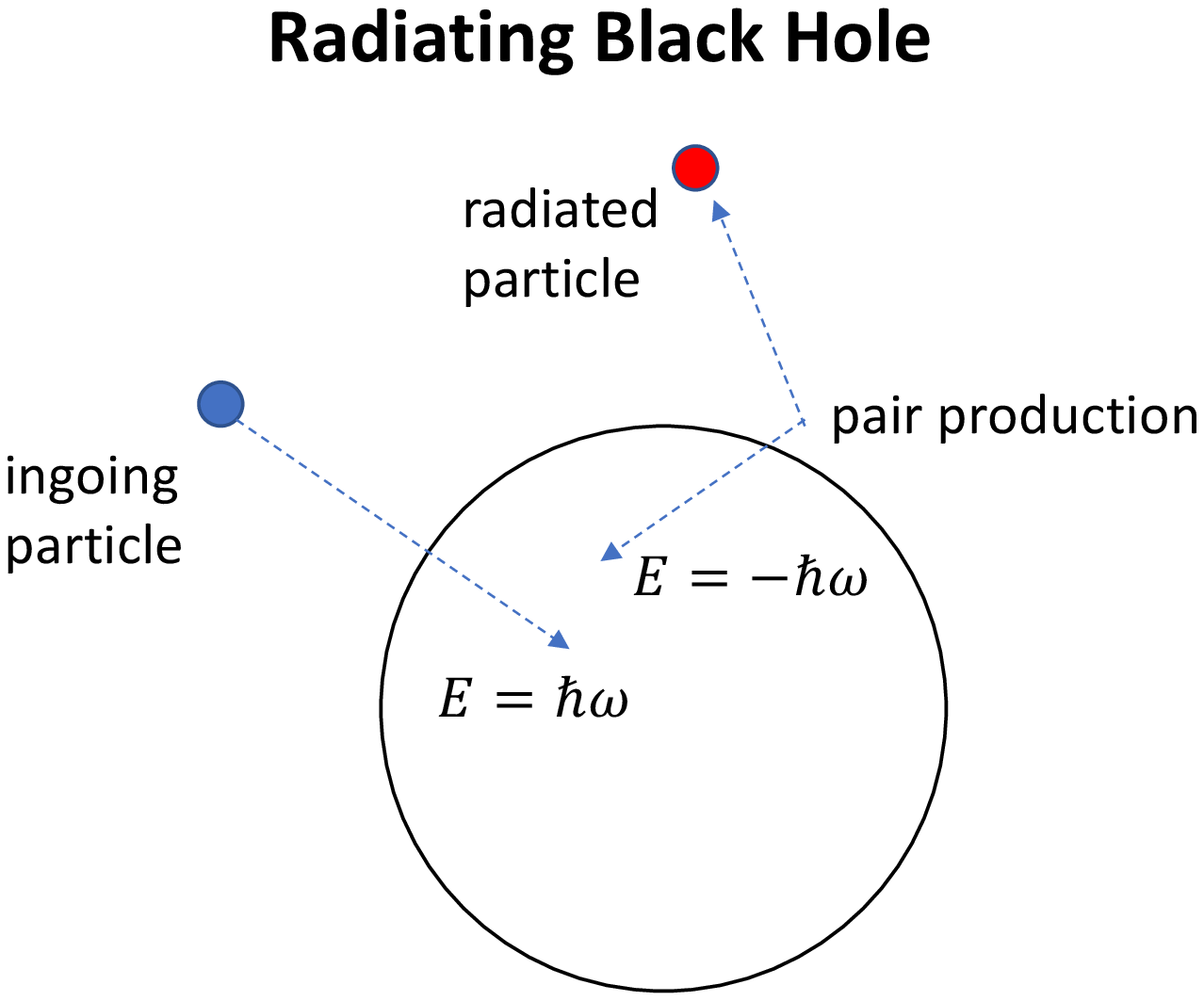}}  \\ \vspace{.2in} 
b)\resizebox{.52\textwidth}{!}{\includegraphics{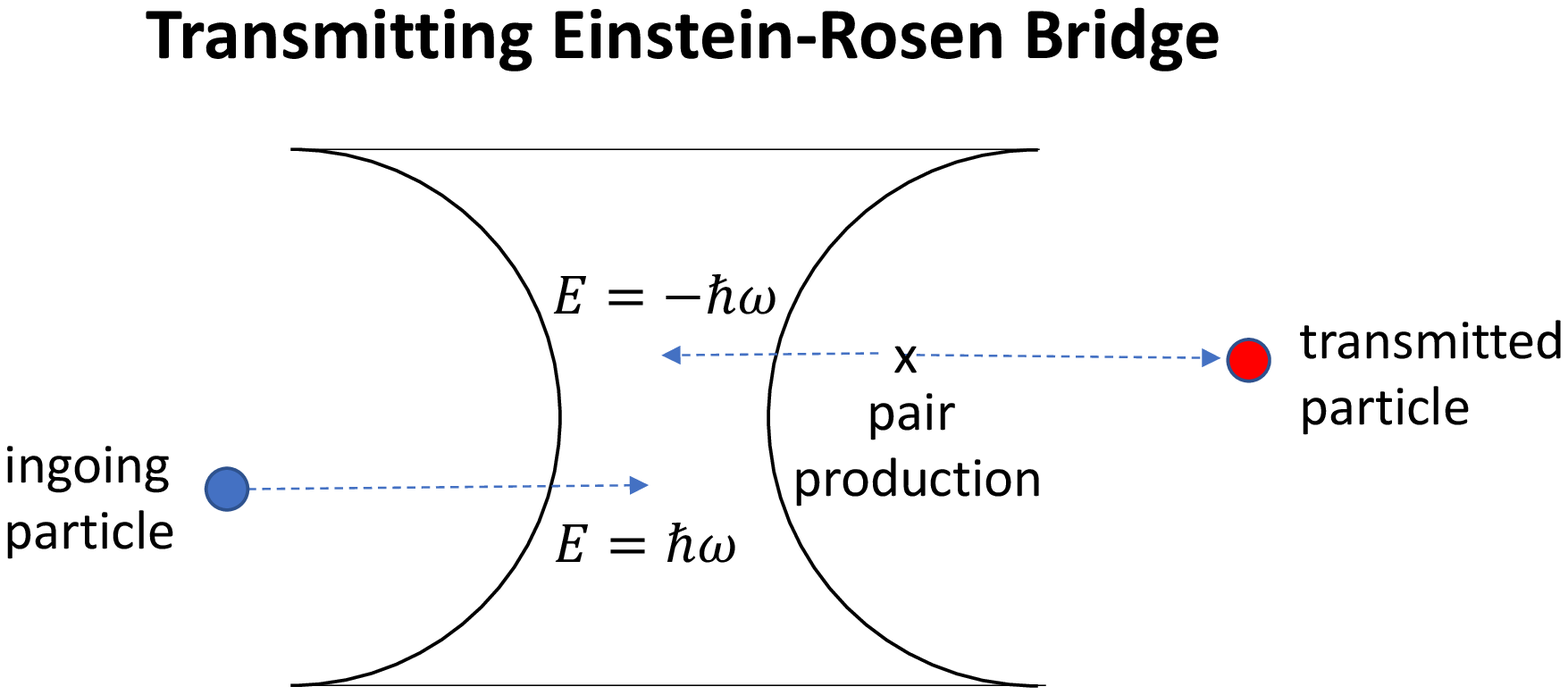}}
 \caption{Steady states of a) a radiating black hole, and b) a transmitting Einstein-Rosen bridge, result from pair production just outside the horizon and ongoing classical ingestion.}
 \label{figRadTrans}
 \end{figure}
 
\section{Background}
\subsection{Quantum potential interpretation of black hole radiation}
If black hole radiation, like other quantum processes, can be described by an objective deterministic theory, it should be possible to augment Einstein's classical equations with terms that give such radiation, which can indeed be characterized as a form of tunneling \cite{ParikhWilczek}. A quantum potential for black hole radiation was in fact derived by deBarros {\it et al.} \cite{deBarrosIJMP,deBarrosPhysLett} in work that has not been widely cited.
Recall that for a simple quantum mechanical system with a wave-function $\Psi$ written in polar form as $\Psi={\cal R} \exp(i{\cal S})$, the quantum potential is $Q=-(\nabla^2 {\cal R})/2m{\cal R}$, meaning that the  Schrodinger equation for $\Psi$ implies that particle motion is governed by the Hamilton-Jacobi equation with an extra potential term:
\begin{equation}
\label{HamJac}
\frac{\partial S}{\partial t} + \frac{(\nabla S)^2}{2 m} + V - \frac{1}{2m}\frac{\nabla^2 {\cal R}}{\cal R}=0
\end{equation}
implying that a single particle follows an augmented equation of motion  $m d^2 x/dt^2 = -\nabla(V+Q)$.  In the common example of tunneling through a potential barrier, we imagine that the particles each  possess  an energy $E < V$, where $V$ is now the height of the barrier. There are no classically allowed trajectories through the barrier, but nevertheless, according to standard quantum mechanics, and as experimentally confirmed, particles that start on one side do appear on the opposite side.  There are no definite trajectories in standard quantum mechanics (although the result could be obtained as a sum over alternative classically forbidden trajectories in the path integral formalism) or in a semiclassical approximation.  But in Bohm's ontological interpretation, each particle follows a definite, classically allowed trajectory in a dynamical system governed by the potential $V+Q$.  That is possible because $E > V+Q$ over the entirety of at least some paths through the barrier.  These paths start from particle positions  in a narrow range of initial values, as illustrated graphically by Bohm and Hiley (Ref. \cite{BohmBook} Section 5.1).  

In the current context, the quantum potential is gleaned from the Wheeler-Dewitt equation $\hat H \Psi=0$, obtained by quantizing the hamiltonian constraint in the ADM formulation of general relativity, in which spacetime is foliated into a temporal sequence of spacelike hypersurfaces.  For a general wave-functional $\Psi$ of the spatial metric on a hypersurface and of the matter fields, this equation is:
\begin{equation}
\label{WdW}
[G_{ijkl} \frac{\delta }{\delta h_{ij}}\frac{\delta }{\delta h_{kl}} -\sqrt{h}\; \RR(\h) + H_{\mbox {matter}}]\Psi(\h,\mbox{matter}) =0
\end{equation}
where  $h$ is the determinant of the space metric $h_{ij}$, $G_{ijkl}\equiv\frac{1}{2}h^{-1/2}(h_{ik}h_{jl} + h_{il}h_{jk} - h_{ij}h_{kl})$ and $\RR$ is the intrinsic curvature of the evolving space-like hypersurface. 
Inserting the polar decomposition  $\Psi={\cal R} \exp(i{\cal S})$ in  (\ref{WdW}), we get:
\begin{eqnarray}
\label{WdWQ}
G_{ijkl} \frac{\delta S}{\delta h_{ij}}\frac{\delta S}{\delta h_{kl}} &-&\sqrt{h}\; \RR(\h) + \sqrt{h}\;Q \nonumber \\
&+& \mbox{matter terms} =0
\end{eqnarray}
for a quantum potential $Q$ that is defined in terms of  derivatives of ${\cal R}$ both with respect to the metric and with respect to the matter fields. For $Q=0$,  Eq. (\ref{WdWQ}) is the Hamilton-Jacobi equation for the usual action $S$ in the ADM formalism. For $Q \ne 0$,  the dynamics are modified, but all fields still follow well-defined trajectories, as for a single particle in the Bohmian interpretation.

Tomimatsu  \cite{Tomimatsu} solved the Wheeler-Dewitt equation (\ref{WdW}) for a spherically symmetric case, using boundary conditions that describe a scalar field in an evaporating black-hole metric, limiting attention to the region near the apparent horizon.  He found one quantum-mechanically interesting solution:
\begin{equation}
\label{eqPsi}
\Psi(r_o,\Phi)=C \exp[i\Big(\frac{r_o}{4}+\frac{k^2}{2r_o}\Big) -|k \Phi|]
\end{equation}
giving the wave-functional $\Psi(r_o,\Phi)$ as a function of the instantaneous apparent black-hole radius $r_o$ and the scalar field $\Phi$ at the apparent horizon, with $k$ an eigenvalue to be determined.  We take the wave functional to have the same form for a transmitting ER bridge, since only the topology is different from that of a black hole. Traversability can indeed be determined by considering the region near the horizon. 
For the wave function (\ref{eqPsi}), we have ${\cal R}=C\exp(-|k \Phi])$, and so the quantum potential depends only on derivatives with respect to the scalar field and not with respect to the metric.  As found by  deBarros {\it et al.}  \cite{deBarrosPhysLett}, the  term in the matter sector of the hamiltonian, $(1/2r_o^2) \hat P_{\Phi}^2 = -(1/2r_o^2) \delta^2 /\delta \Phi^2$, contributes a quantum potential  
\begin{equation}
\label{Qpot}
Q= -\frac{k^2}{2 r_o^2}
\end{equation}
to the Hamilton-Jacobi equation (\ref{WdWQ}). The effect of the added term is to pull the black hole radius $r_o$ toward zero.
We assume here that the quantum evaporation/transmission process induced by the quantum potential (\ref{Qpot}) is balanced by a purely classical accretion/particle-entry process to maintain a steady state (Fig. \ref{figRadTrans}). 

\subsection{Classical higher-derivative gravity}

In this paper, we seek a generally covariant extension of Einstein's equations that gives a quantum potential of the form (\ref{Qpot}) in the corresponding generalization of the Wheeler-Dewitt equation for a radiating black hole in a steady state.   A simple candidate  is constructed by adding terms containing higher derivatives of the metric to the standard second-order equations.  While such higher-derivative extensions of general relativity have become familiar in the context of quantum corrections to general relativity, here we regard the resulting theory simply as an alternative to the standard classical theory.

A textbook derivation of Einstein's equations, e.g. Weinberg \cite{Weinberg} Sect. 7.1, relies not only on general covariance, but on an explicit assumption that the equations are second order, or equivalently, that they are scale invariant.  General relativity can in fact be extended to theories of the form: 
\begin{equation}
\label{modGR}
R_{\mu \nu} -\frac{1}{2}Rg_{\mu \nu} + g_{\mu \nu}\Lambda 
+ \sum_{n>2} c_n L^{n-2} R^{(n)}_{\mu \nu} = 8\pi T_{\mu \nu}
\end{equation}
where $R^{(n)}_{\mu \nu}$ is a quantity involving a total of $n$ derivatives of the
metric, $L$ is a fundamental length scale, the $c_n$ are dimensionless
constants and we have included a cosmological constant $\Lambda$ for full generality. If $L=L_P$, the Planck length, then the new
terms in the extended theory (\ref{modGR}) are negligible on macroscopic scales. They need only be considered
if curvature is significant at the Planck length scale. It might be hoped that no macroscopic effects would ensue in that case, but this is not guaranteed because of the nonlinearities.

A fourth-order classical theory was introduced by Stelle \cite{Stelle77}, in order to create a quantized version that is renormalizable.  However, the resulting quantum theory is plagued by ``ghosts", or negative-norm states, that violate unitarity.  The classical theory, whose main weakness is instability related to the ghosts in the quantum version, was explored in \cite{Stelle78}.  A 4th-order classical theory, intended for use as such,  was studied by Ruzmaikina and Ruzmaikin \cite{Ruz} for application to cosmology. More recently, a thorough study of 4th order gravity by Lu  {\it et al.} \cite{Lu, LuPRL}, which we shall rely on heavily, was enabled by the advent of algebraic calculation tools like Mathematica.  Though Lu {\it et al.} stated that the principal intended application was to compute quantum corrections, their main result is a compendium of classical, spherically symmetric, time-independent solutions.

\section{Equivalent of the quantum potential in 4th order gravity} 

A generally covariant description of the quantum potential term must exist to satisfy the principle of relativity.  The same geometrodynamics and the same quantum potential will describe both standard black-hole evaporation and a steady state of the black hole or bridge.  So we seek  extensions of Einstein's equations that would produce the quantum potential term classically, starting with 4th order extensions.
As with a potential barrier in simple quantum mechanics, the only way that the resulting classical equations would permit ``tunneling" through a modified ER bridge is for the modifications  to make the bridge formally traversable.  That is indeed what we will find, even though the quantized theory and resulting quantum potential were constructed to describe behavior only near the horizon. 

For pure gravity, the most general 4th order extension of Einsteinian general relativity \cite{Lu} is given by a Lagrangian density of the form
${\cal L}=\sqrt{-g}[\gamma R - \alpha C_{\mu \nu \rho \sigma} C^{\mu \nu \rho \sigma} + \beta R^2]$ where $C_{\mu \nu \rho \sigma}$ is the Weyl tensor, formed by removing non-vanishing contractions from the curvature tensor $R_{\mu \nu \rho \sigma}$.  This Lagrangian density can also be written:
\begin{equation}
\label{L4}
{\cal L}=\sqrt{-g}\;[\gamma R - 2\alpha R_{\mu \nu} R^{\mu \nu} + (\beta + \frac {2\alpha}{3}) R^2]
\end{equation}
by using the topological invariance of a quantity specified in the Gauss-Bonnet theorem \cite{Lu}.

An  ADM-type formulation of 4th order gravity  is derived as for ordinary general relativity. The Gauss-Codazzi relations \cite{GaussCod, ADM} are used to express the Riemann tensor, the Ricci tensor and the Ricci scalar in terms of the corresponding objects for 3-dimensional space-like hypersurfaces, denoted $\RR_{\mu \nu \sigma \rho}$ etc.,  and the extrinsic curvature $K_{ab}$ of those hypersurfaces, together with lapse and shift variables that describe the flow of time and the point-to-point connections, respectively, between hypersurfaces.
The canonical momentum $p_{ab}$ that is conjugate to $h_{ab}$ is $p_{ab}={\partial{\cal L}}/{\partial \dot h_{ab}}$.
The Hamiltonian density is constructed from ${\cal H}(\p,\q) =  \p \cdot \dot\q -{\cal L}(\q, \dot\q)$, with $\q=\h$, by writing the kinematic term
$ \p \cdot \dot\h$ also in terms of the extrinsic curvature as with ordinary gravity, and expressing the extrinsic curvature in terms of the canonical momenta  \cite{ADM}.  The lapse appears as a Lagrange multiplier of the hamiltonian constraint, as in the usual ADM approach.  Let us write the hamiltonian constraint in the form:
\begin{eqnarray}
\label{Ham3}
F(\p,\dot \p,\ddot\p) -\sqrt{h}\;\big[\gamma \RR &-& 2 \alpha \RR_{a b}\! \RR^{a b} \nonumber \\
                                                    &+& (\beta + \frac {2\alpha}{3})\;(\RR)^2\big] + {\cal H}_{matter}=0 
\end{eqnarray}
where the function $F(\p,\dot\p,\ddot\p)$ depends also on curvature, to isolate the \p-independent terms.  (Greek indices assume values in $\{0,1,2,3\}$ while latin indices are restricted to $\{1,2,3\}$.)  \footnote{The function $F(\p,\dotp,\ddotp)$ for the case $\alpha=0$ is given explicitly in terms of extrinsic curvature in Ref. \protect\cite{Cotsakis}.}

This constraint is to be compared to the one for ordinary gravity, with quantum potential added, which is
\begin{equation}
\label{ADM2Q}
G_{ijkl} \;p^{ij} p^{kl} - \sqrt{h}\;\gamma  \RR - \sqrt{h}\; \frac{k^2}{2 r_o^2} + {\cal H}_{matter}=0
\end{equation}
from which equation (\ref{WdWQ}) is obtained by substituting $p_{ab}\rightarrow \partial S/\partial h_{ab}$.
Eq. (\ref{Ham3}) is generally 4th-order in the canonical momenta while the corresponding equation for ordinary gravity (\ref{ADM2Q}) is 2nd order, but we claim that we can ignore the terms in $\p$. 
That is because for static metrics such as we will consider here, hypersurfaces of constant time have zero extrinsic curvature $K_{ab}\equiv \nabla_a n_b=0$, where $\n$ is a time-like unit vector field normal to the surface, and $\nabla$ is the projection of the covariant derivative onto the hypersurface.  For such metrics we can take $\n(\x)=(n_t,0,0,0)$.  The Lagrangian density ${\cal L}$ can be written as a sum of  terms with factors of the intrinsic curvature $\RR_{ab}$ or its contraction $\RR$, and terms with factors that are quadratic in the extrinsic curvature $K_{ab}$ or that are  time derivatives of  $K_{ab}$, since the metric is Gaussian normal (Ref. \cite{ADM} Sect. 21.5) as restricted to the horizon.  Using    ${\partial \RR_{ij}}/{\partial \dot h_{ab}}=0$ and ${\partial K_{ab}}/{\partial \dot h_{ab}}=1/2 l$, where $l$ is the time-independent lapse, and  $K_{ab}=0$, we find that  $p_{ab}={\partial{\cal L}}/{\partial \dot h_{ab}}=0$.  So the kinematic term $\p \cdot \dot\h$ vanishes and since $K_{ab}=0$, the Lagrangian contributes only two new terms, in $ \RR_{a b}\! \RR^{a b}$ and  $(\RR)^2$, and so $F=0$.  The steady states of the dynamics defined by the Hamiltonian (including the matter terms), as would include ER bridges or wormholes prior to collapse, are the same  in quantized ordinary general relativity and in classical 4th order gravity, if near the horizon:
\begin{eqnarray}
\label{GR4QGR}
-(\gamma - \gamma_{\scriptscriptstyle {GR}}) \RR &+& 2\alpha \RR_{a b}\!  \RR^{a b}   \nonumber \\
                             &-& (\beta + 2 \alpha/3)( \RR)^2 =-\frac{k^2}{2 r_o^2}
\end{eqnarray}
where $\gamma_{\scriptscriptstyle {GR}}=1/16\pi G$ is the canonical value of $\gamma$. (The usual ``supermomentum" part of the total Hamiltonian vanishes for $\p=0$.) 
It is noted that the negative sign of the quantum potential $Q$, viewed as a contribution to the energy density at the horizon, is what is required of the ``exotic matter" that could serve to violate the averaged weak energy condition and to keep a traversable wormhole open  \cite{AWEC}.  

For black holes and Einstein-Rosen bridges, we generally have $R_{\mu \nu}, R = O(1/r_o^2)$.  The requirement of Eq. (\ref{GR4QGR}) is thus that that the $1/r_o^4$ terms in the
expressions that are quadratic in curvature,  $\RR_{a b}\!  \RR^{a b} $ and $( \RR)^2$ cancel, and that these quadratic curvature expressions, taken together, are  $O(1/r_o^2)$.  We consider spherically symmetric, time-independent metrics of the form:
\begin{equation}
\label{metric}
ds^2= -B(r)dt^2 + A(r) dr^2 + r^2 d\theta^2 + r^2 \!\sin^2 \theta \;d\phi^2
\end{equation}
with the  spatial metric on hypersurfaces of constant $t$ given trivially by: $h_{ij}=g_{ij}$.  For a metric of this form,  the indices $0,1,2,3$ are more specifically written as $t,r,\theta,\phi$, respectively. The Ricci tensor on the hypersurfaces is diagonal with three non-vanishing elements:
\begin{eqnarray}
\label{Rdiag}
\RR_{rr}&=& -\frac{1}{r} \frac{A'}{A}    \nonumber \\
\RR_{\theta\theta}&=&-1-\frac{A'}{2A^2}+\frac{1}{A}\\
\RR_{\phi\phi}&=& \sin^2\theta\left(-1-\frac{A'}{2A^2}+\frac{1}{A}\right)\nonumber
\end{eqnarray}
where primes denote derivatives with respect to $r$. Denoting $A^{-1}(r_o) \RR_{rr}(r_o)\equiv a$,  $r_o^{-2}\RR_{\theta\theta}(r_o)\equiv b$, and  $r_o^{-2}\sin^{-2}\!\theta \;\RR_{\phi\phi}(r_o)\equiv c$, we have at the horizon:
\begin{equation}
\label{Rabc}
\RR=a + b + c
\end{equation}
\begin{equation}
\label{Rijabc}
\RR_{ij}\RR^{ij}=a^2 + b^2 + c^2
\end{equation}
Assuming $a,b,c \sim 1/r_o^2$, the requirement that the quadratic curvature terms mimic the quantum potential is that the terms of leading order in $1/r_o$ vanish, and that:
\begin{eqnarray}
\label{zeroquart}
2 \alpha \RR_{a b}\! \RR^{a b}- (\beta + \frac {2\alpha}{3})\;(\RR)^2&=& \nonumber \\
2 \alpha(a^2 + b^2 + c^2)-(\beta + 2\alpha/3)(a+b+c)^2&=& \nonumber \\
\left(\frac{4 \alpha}{3} - \beta\right)a^2 -4\left(\beta- \frac{ \alpha}{3}\right)b^2 - 4\left(\beta+\frac{2 \alpha}{3}\right)ab&=& \nonumber \\
\frac{-k^2}{2r_o^2}
\end{eqnarray}
having noticed that $b=c$, and having taken $\gamma=\gamma_{\scriptscriptstyle {GR}}$, as we will henceforth for agreement with general relativity at large scales.

We have solved for the metric (\ref{metric}) in terms of $A(r)$ and $B(r)$, in series expansions, following the method of Lu {\it et al.} \cite{Lu}:
\begin{subequations}
\label{seriesAB}
\begin{eqnarray}
\frac{1}{A(r)}\equiv f(r)&=&f_1(r-r_o)+f_2(r-r_o)^2+\ldots\\
B(r)&=&b_o + b_1(r-r_o) + b_2(r-r_o)^2+\ldots \nonumber\\
\end{eqnarray}
\end{subequations}
There are indeed wormhole solutions with $A(r) = \infty$  at the horizon $r=r_o$, as with a  Schwarzchild black hole, but $B(r_o) \ne 0$, unlike a black hole in ordinary gravity or in 4th order gravity \cite{LuPRL}.   Such traversable wormholes were studied by Lu {\it et al.} \cite{Lu}  in the restricted case $\beta=0$.  Here, $\beta \ne 0$ is needed so that there are non-trivial solutions to
(\ref{zeroquart}) with $\alpha \ne 0$ and different from ordinary gravity. The 4th order theory admits traversable wormhole solutions, as in \cite{Lu}, but with more free parameters than for the $\beta=0$ case.  Quantum tunneling is thus accounted for classically.  

The equations of motion are derived by minimizing the action $I=\int d^4x\sqrt{-g}\; {\cal L}$ formed from the Lagrangian density (\ref{L4}). They are:
\begin{eqnarray}
\label{eqmotion}
0&=&\frac{1}{\sqrt{-g}} \frac{\delta I}{\delta g^{\mu \nu}}\equiv H^{\mu \nu}  \nonumber \\
   &=&-\gamma(R_{\mu \nu} - \frac{1}{2}g_{\mu \nu}R) - \frac{2}{3}\left(\alpha-3\beta\right)R_{;\mu ;\nu} \nonumber \\
   &&+2\alpha g^{\rho \sigma}R_{\mu \nu ;\rho ;\sigma}  - \frac{1}{3}\left(\alpha+6\beta\right)g_{\mu \nu}g^{\rho \sigma}R_{;\rho ;\sigma} \nonumber \\
   &&+ 4\alpha R^{\rho\sigma}R_{\mu\rho\nu\sigma} + 2\left(\beta + \frac{2}{3}\alpha\right)R R_{\mu\nu} \nonumber \\
   &&+ \alpha g_{\mu\nu}R^{\rho\sigma}R_{\rho\sigma} - \frac{1}{2}g_{\mu\nu}\left(\beta + \frac{2}{3}\alpha\right) R^2
\end{eqnarray}
where a semicolon denotes covariant differentiation with respect to the index following.  One first substitutes the spherically symmetric, static metric form (\ref{metric}) in the usual expressions for the Ricci tensor (see e.g. \cite{Weinberg}), to get 4-dimensional expressions for $R_{\mu \nu}$ that are extensions of (\ref{Rdiag}), and inserts these expressions in (\ref{eqmotion}) to derive differential equations for $A(r)$ and $B(r)$.  Combining the equations for $H^{tt}$ and $H^{rr}$ (sufficient to determine $H^{\theta\theta}$ and $H^{\phi\phi}$ as well, because of a Bianchi identity) in the manner of  Ref. \cite{Lu}, one derives a pair of third-order differential equations for $A(r)$ and $B(r)$ that are found to match those given in the Appendix to \cite{Lu}.  The series expansions (\ref{seriesAB}) were substituted in those equations and solved for the lowest order coefficients in Mathematica (see Supplementary Material).  We find:
\begin{subequations}
\label{lowcoeffs}
\begin{eqnarray}
f_1&=&\frac{1}{3}\sqrt{\frac{-2(\alpha-3\beta)(2\alpha -6\beta - 3 \gamma r_o^2)}{3\alpha \beta r_o^2}}\\
b_1/b_0&=&\frac{2(\alpha + 6 \beta)}{(\alpha -3\beta)r_o}
\end{eqnarray}
\end{subequations}
The condition $b_0 > 0$ implies that the solution is a wormhole,  as in \cite{Lu}. But unlike the $\beta=0$ case, we find that $b_2$ is unconstrained, and therefore can be tuned for asymptotic flatness. It can be checked that the extra terms in (\ref{eqmotion}), as compared to the usual second-order Einstein equations,
give an effective negative energy contribution, when averaged along a null geodesic through the interior of the wormhole, that violates the averaged weak energy condition, as required for traversability.  

To satisfy the equivalence condition (\ref{zeroquart}), we first note that the curvature variables $a=A^{-1}(r_o) \RR_{rr}(r_o)$ and $b=c=r_o^{-2}\RR_{\theta\theta}(r_o)$, where   $\RR_{i j}$ is given by (\ref{Rdiag}),  depend only on the leading coefficient $f_1$:
\begin{eqnarray}
\label{ab}  
a&=&\frac{f_1}{r_o}  \nonumber   \\
b=c&=&-\frac{1}{r_o^2} + \frac{f_1}{2r_o}
\end{eqnarray}
We will use the solution (\ref{lowcoeffs}a) for $f_1$ in terms of $\alpha$ and $\beta$ to compute the quadratic curvature contribution (\ref{zeroquart}), which we now denote $V(r_o)$. 
\begin{eqnarray}
\label{Vr0}
V(r_o)&\equiv&\left(\frac{4 \alpha}{3} - \beta\right)(a(\alpha,\beta))^2 - 4\left(\beta- \frac {\alpha}{3}\right)(b(\alpha,\beta))^2    \nonumber \\
          &&- 4\left(\beta+\frac{2 \alpha}{3}\right)a(\alpha,\beta)b(\alpha,\beta)
\end{eqnarray}
Substituting from (\ref{ab}) and (\ref{lowcoeffs}), the condition for equivalence of  $V(r_o)$ to the de Barros {\it et al.} quantum potential becomes:  
\newcommand{\gammaGR}{\gamma_{\scriptscriptstyle {GR}}} 
\begin{eqnarray}
\label{Vr0Q}
V(r_o) &\equiv& \frac{2}{81 r_o^4 \omega}\Bigg\{3\gammaGR r_o^2(36-15\omega + \omega^2)   \nonumber \\
          &+&\beta\Bigg[216- 2 \omega^3 \nonumber \\
          &+& 36 \omega\left(-9 + \sqrt{\frac{6(-3+\omega)(6\beta+3\gammaGR r_o^2 - 2 \beta\omega)}{\beta\omega}}\right) \nonumber \\        
         &+&6\omega^2\left(15+ \sqrt{\frac{6(-3+\omega)(6\beta+3\gammaGR r_o^2 - 2 \beta\omega)}{\beta\omega}}\right)\Bigg]\Bigg\} \nonumber \\
         &= &-\frac{k^2}{(2 r_o^2)}
\end{eqnarray}
where we have expressed $V$ in terms of $\beta$ and $\omega\equiv\alpha/\beta$ for convenience.

\begin{figure}
a)\resizebox{.37\textwidth}{!}{\includegraphics{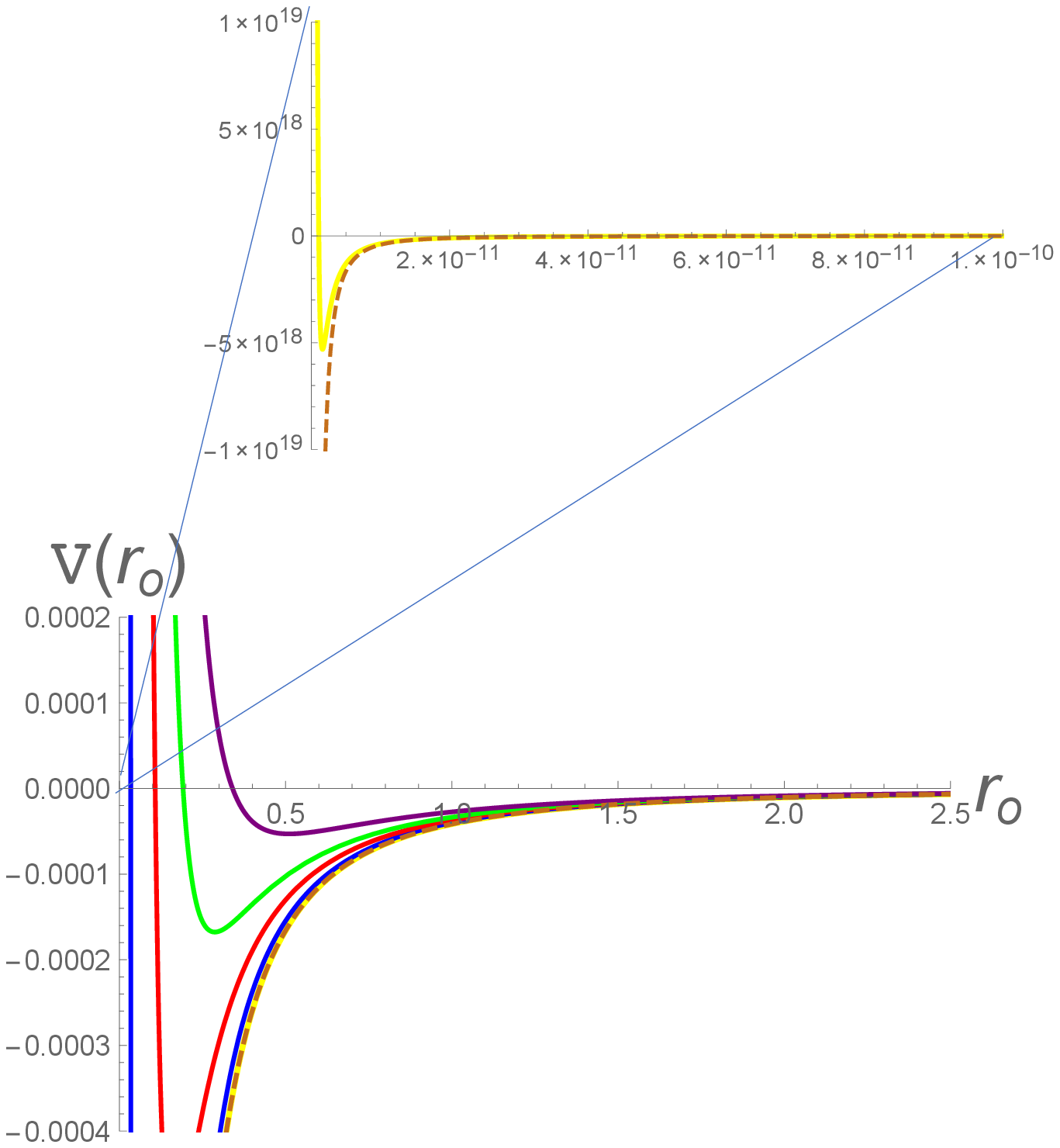}}  \\ \vspace{.2in} 
b)\resizebox{.37\textwidth}{!}{\includegraphics{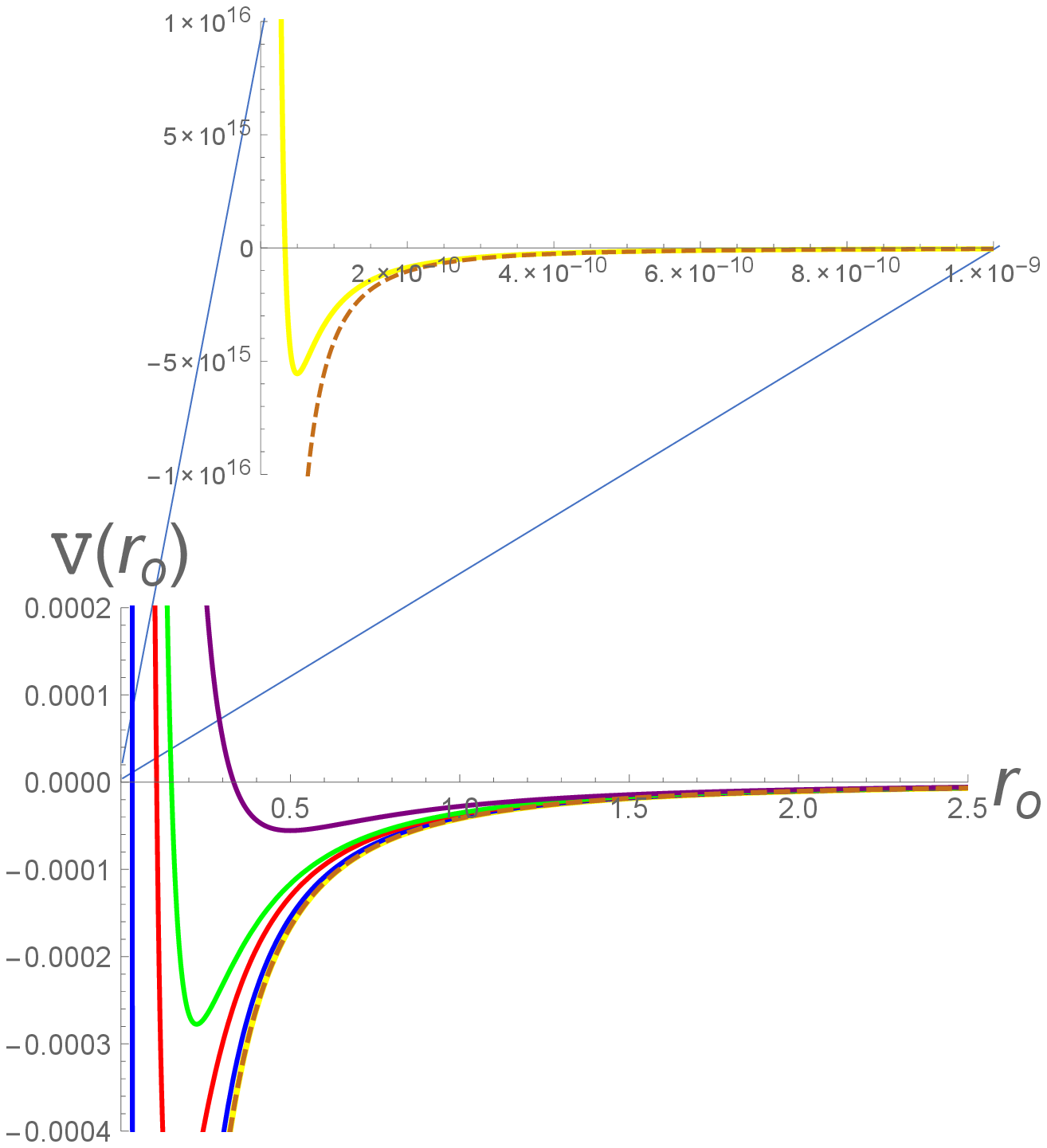}}
 \caption{The contribution of the quadratic curvature terms to the hamiltonian constraint, vs. wormhole radius $r_o$, for a) decreasing values of $\beta$: $\beta= 10^{-7}$ (purple), $10^{-7.5}$ (green), $10^{-8}$ (red), $10^{-9}$ (blue), and $10^{-30}$ (yellow), while the the ratio of the coefficients is held fixed at the first solution to (\ref{eqratio}), $\alpha/\beta \approx 3.009$; and for b) values of
$\beta$: $\beta= 10^{-10}$ (purple), $2\times 10^{-11}$ (green), $10^{-11}$ (red), $10^{-12}$ (blue), and $10^{-30}$ (yellow), while the ratio is held at the second solution, $\alpha/\beta \approx 11.962$. The quadratic curvature contribution $V$ converges to the quantum potential $-k^2/(2 r_o^2)$ (dashed line) as $\alpha,\beta \rightarrow 0$ in both
cases, except for large positive values for a decreasing range of small $r_o$, as shown for the smallest value $\beta=10^{-30}$ (yellow line) in expanded views about the origin.}
 \label{figV}
 \end{figure}

The coefficient of the second term in the braced expression, $\beta$, must vanish to avoid a $1/r_o^4$ contribution for small $r_o$.  Then the first term gives
\begin{equation}
\label{eqratio} 
2\gammaGR(36-15\omega + \omega^2))/(27\omega) = -\frac{k^2}{2}
\end{equation} 
a quadratic equation that can be solved for $\omega$, for given $k$. Since the same geometrodynamics must describe black-hole evaporation,  a value of the eigenvalue $k$ can be obtained by matching the black-hole evaporation rate as computed by deBarros {\it et al.} \cite{deBarrosPhysLett}, $k^2/4M^2$, to the Hawking rate  \cite{Hawking} $P= \hbar c^6/(15360 \pi G^2 M^2)$. This gives $k=\sqrt{\hbar c^6/3840 \pi G^2}$, or in units for which $\hbar=c=G=1$, $k \approx 0.01$. 
 In the limit $\alpha,\beta \rightarrow 0$, the desired value of the ratio of the coefficients $\omega= \alpha/\beta$ solves (\ref{eqratio}) and is found to be $\alpha/\beta=\omega=3.009$ or $\alpha/\beta=11.962$. For $\alpha/\beta$ fixed at these values, with $\gammaGR=1/16\pi$, the contribution of the quadratic curvature terms is plotted vs. $r_o$ in  Fig. \ref{figV} as the two coefficients each approach zero.  As we approach the limit, the quadratic curvature $V$ contribution matches the quantum potential  $-k^2/(2r_o^2)$   down to smaller and smaller radii $r_o$, but very large values of $V$ are found for a  narrowing range around $r_o=0$. We note that the large potentials could conceivably help in avoiding divergences due to an arbitrarily large tendency to form arbitrarily narrow wormholes. Otherwise the quantum potential can be matched as closely as desired for sufficiently small coefficients.
 
\section{Discussion and Conclusions} 

\subsection{The nature of non-locality}
Most of the non-locality implied by Bell's Theorem and featured in the Bohm-deBroglie interpretation is already contained in the ER=EPR conjecture. The required extension is only slight.  Maldacena and Susskind's conjecture  started from macroscopic black holes and bridges, for which quantum tunneling is of less direct importance.
Alice, at one end of such a bridge, can send any kind of object or signal through the bridge (including a firewall) that Bob will encounter as soon as he crosses the horizon from the other end.
But macroscopic objects cannot reach him before that time, either classically or through tunneling, just as they cannot tunnel through a potential barrier in simple quantum mechanics. The impact of tunneling suggested here is for Planck-scale configurations, where it is enough to take us from the Copenhagen interpretation to the ontological one, in one sense a slight change.

With tunneling, a mechanism for the violation of Bell's inequality is conceivable, provided that the results derived here extend to vector fields propagating through an ``ER"  bridge joining a pair of spinor particles. In this regard, it is noteworthy that wormhole solutions with electric charge in a quadratic Palatini theory that includes small quadratic curvature terms with a fixed ratio have been previously reported \cite{Olmo}.  With such a connection, the action of a magnetic field on one member of
an EPR pair in a spin measurement  could instantaneously affect the electromagnetic signal leaving the bridge that determines the measured spin of the partner.  The orientation of the first magnetic field is information that could be tunneled, violating an assumption used in deriving Bell's inequality - that the second measurement is independent of the first orientation.

Tunneling is of course possible in ordinary quantum mechanics or semiclassical approximation thereof, without the Bohm interpretation. But the usual (Copenhagen) interpretation of tunneling is plagued by the same peculiar features of quantum theory, including non-locality, that we seek to explain. In the Copenhagen interpretation, there are no definite trajectories, so an account of non-locality  in terms of extended general relativity, itself a deterministic theory, would not be possible.

In the higher-derivative gravity account of non-locality particles can follow definite trajectories through the wormhole.  So one must explain why such behavior could not be used to 
transmit super-luminal signals.   We rely on the same argument given by Bohm and Hiley  (Ref. \cite{BohmBook} Sections 7.2, 12.6) to negate this possibility: In the absence of a Maxwell's Demon, one cannot fix all the hidden variables controlling the spin of an individual particle, and so must examine a classical ensemble of states. The statistics of the ensemble on the opposite side  of the bridge can be shown to be unaffected by the measurement.  To fix enough of the hidden variables to transmit information, a Demon would have to violate the uncertainty principle, which is preserved in Bohm's interpretation.

\subsection{Measurement}
The process of measurement has not been described here, since the focus of our analysis has been on steady entangled states. Collapse of the wormhole has not been discussed.  It is thought that the fourth-order time derivatives in the classical hamiltonian constraint (\ref{Ham3}), away from the steady state, could describe a physical process of wormhole/wavefunction collapse on very short time scales. We note that the instabilities inherent in the 4-th order theory, the ones that give rise to ghosts in a quantum version, could play a role. Instability due to the absence of a lower bound on the Hamiltonian follows from a general result on the instability of any classical theory (even a theory of a single particle) given by a Lagrangian that is a function of higher time-derivatives of the canonical coordinates  \cite{Ost}.  While we cannot give a full account of the effects of such instability - and that is the major weakness of higher-derivative gravity theories generally - we suggest that this instability need not be fatal.  For a time-independent solution, the absence of a lower energy bound is only relevant when the system interacts.  Interactions conserve total energy and in simple cases need not lower the gravitational energy at all.  The relevance is statistical. A large number of random interactions are nearly certain to lower the gravitational energy, and raise the energy in other sectors, without bound  \cite{Ost}.  
For interactions such as that with a measuring apparatus, it seems natural to refer to Susskind's notion of the transfer of entanglement when an ER-bridge interacts with its environment \cite{SusskindEverett}.  The implied proliferation of wormholes, each containing a region of negative gravitational energy, appears to be consistent with the unboundedness of the gravitational Hamiltonian.

If entanglement between an EPR pair is broken by measurement, as in the usual interpretation, we imagine that the entry of a single particle triggers collapse, preventing \cite{DuaneEntropy} the problems due to closed time-like curves through macroscopic  wormholes and the resulting  vacuum polarization divergence, due to quanta re-circulating on nearly closed spirals, that was previously debated between Kim and Thorne \cite{KimThorne} and Hawking \cite{HawkingVsKT}.  For a Planck scale wormhole, the collapse itself (with sensitivity to the orientation of the remote measuring field, as required by Bell's theorem) could indeed be the only occasion on which effects are transmitted.  But one could not control an individual collapse so as to transmit information.

\subsection{Implications for general relativity and quantum theory}
The large-scale predictions of general relativity could also be expected to remain intact.  For black holes or non-traversable ER bridges of macroscopic size, the quadratic curvature terms in the above analysis have no significant effect on the dynamics  that is different from that of the quantum potential, or therefore, of the usual quantum theory. That is, at large scales, the effect of the new terms is well approximated by the (very small) quantum potential $Q=-k^2/(2r_o^2)$ that was originally put forward to explain macroscopic black hole evaporation \cite{deBarrosPhysLett}.  (Note that the width of the wormholes is not determined in the above analysis and is not the macroscopic radius $r_o$.) Furthermore, the potential that follows from the quadratic curvature terms is unique, and does not depend on a choice of eigenstate as in the treatment of Tomimatsu \cite{Tomimatsu} where classical (non-radiating) black hole behavior occurs for another wave function solution.  So the ambiguity  reported by deBarros {\it et al.} \cite{deBarrosPhysLett} regarding black hole evaporation in a quantum potential treatment, is avoided if the potential comes from extra classical curvature terms. An absence of macroscopic effect might have been expected if the coefficients $\alpha$ and $\beta$ (which have the dimensions ${\mbox{ [length]}}^2$) had Planck-scale values. That these coefficients can be even smaller, indeed infinitesimal, strengthens the argument that the macroscopic predictions of ordinary general relativity are preserved.

In the interpretation of black hole evaporation, it is not necessary, as might be feared,  to replace the macroscopic quantum black hole with a single classical wormhole-like object of macroscopic size. Nor should we imagine that such a traversable object exists in the case of entangled macroscopic black holes. We can instead posit a collection of Planck-scale wormholes for individual particles that would piece together to solve the 4th-order equations.

While we do not want large deviations from the observed consequences of general relativity, there should be testable consequences of the theory proposed here. At sub-Planck scales, where $r_o<<L_P$,  the potential arising from the quadratic curvature terms can deviate strongly from that predicted by quantum theory, for any values of $\alpha$ and $\beta$ that remain finite.  Differences could have detectable physical consequences.  The higher-derivative terms would be large in the very early universe, with possible implications for cosmology.

Secondly, we note that the quantum potential depends on specific combinations of the matter fields with gravity. The quantum potential (\ref{Qpot}) used here arises from a single scalar field in the matter portion of the hamiltonian. Other forms would obtain for vector or spinor fields, such as considered in the charged case \cite{Olmo} mentioned above.  It is possible that allowable combinations of matter fields are subject to a supersymmetric constraint  required to maintain the equivalence between the quantum potential due to the matter fields and contributions from the extra curvature terms.  The prescribed combinations could then be compared with the known spectrum of particles.

Gravity need not be quantized to explain black hole evaporation or tunneling though ER bridges, since Hawking's original account relied only on quantum mechanics in curved space-time.\cite{Hawking}. 
Accordingly this study tells us little about how to quantize gravity, except to assert that micro-wormholes play a role.  On the other hand, it is significant that our quantum potential was obtained from the quantum-mechanical wave functional (\ref{eqPsi})  of both the gravitational field (through the black hole radius $r_0$) and the matter fields, solving the Wheeler-DeWitt equation. Thus there seems to be advantage in treating gravity on the same footing as the matter fields. 
 Our resulting 4th-order theory happens to be renormalizable.  
A previous attempt to quantize gravity, using the asymptotic safety (AS) approach \cite{AS} to defining renormalizable equations with higher derivatives, with the intent of avoiding ghosts, has yielded similar results: 4th order equations involving very small coefficients with fixed ratios and a steady state black hole solution \cite{Cai}.  In the absence of a consistent quantum theory of gravity, we have preferred to discuss the classical instability problem,  as above in connection with wormhole collapse, rather than to examine its quantum counterpart in the problem of ghosts.  But the origin of the quantum potential in Wheeler-DeWitt, the renormalizability, and the similarity to the AS results suggest that the entanglement mechanism proposed in this paper has a bearing on the quantization of gravity that remains to be elucidated.

It is tempting to elevate the proposed role of traversable wormholes in mediating entanglement to a full interpretation of quantum mechanics.  It would first be necessary to explain why the wormholes are of width O(1) (in non-dimensional units), from which Planck's constant would emerge.   One possible source of the restriction is just the weak vacuum recirculation divergence \cite{KimThorne,HawkingVsKT} that precludes macroscopic widths, but a thorough analysis is needed.
Then further, in addition to the need for a description of collapse, one would need to show that all of quantum theory follows from nonlocal entanglement conjoined with some properly constructed, local classical theory.  At Planck scale, that classical theory could be given by adding a matter component of as yet unknown form to our fourth-order gravity Lagrangian, but in a space-time with topology that is not pre-determined. At intermediate scales, there would be effective Lagrangians of form also as yet unknown.

While nonlocal entanglement is not everything, it is certainly central. It is hypothesized that at least the non-local part of the quantum potential, in the general situation, arises from the multiply connected topology. As with an Einstein-Rosen bridge, entangled particles connected by a  traversable wormhole with $r_o$ of Planck scale are not separated within the wormhole and are in a sense the same particle, until the connecting geometry collapses.  The resulting negation of the usual concept of the continuum establishes a new order in the physical world, best described in an atomized spacetime together with features such as non-standard metrics in state space \cite{Palmer} and an ontology given by non-locally defined basis states  that are not easily specified \cite{tHooft}. The view taken here is that an account of how the new order forms from the breakdown of the classical order through gravitational instability of the populated vacuum at micro-scale, could ultimately realize Einstein's vision of a more complete, more detailed and more  predictive version of quantum theory.

{\bf Acknowledgments}

 This work was partly supported by European Commission Grant 658602.




\end{document}